\documentclass[aps,pra,twocolumn,showpacs,amsmath,floatfix]{revtex4}

\usepackage{bm}
\usepackage{epsfig}

\newcommand{\eab}{\varepsilon_{\alpha\beta}}
\newcommand{\ab}{{\alpha\beta}}
\renewcommand{\d}{\partial}

\begin{document}

\title{Virial theorems for vortex states in a confined
Bose-Einstein condensate}
\author{N. Papanicolaou}
\affiliation{
Department of Physics, University of Crete, and Research Center of Crete,
Heraklion, Greece}
\author{S. Komineas and N.R. Cooper}
\affiliation{
Theory of Condensed Matter Group,  Cavendish Laboratory,
Madingley Road, Cambridge CB3 0HE, United Kingdom}

\date{\today}

\begin{abstract}
We derive a class of virial theorems which provide stringent
tests of both analytical and numerical calculations of vortex
states in a confined Bose-Einstein condensate. In the special
case of harmonic confinement we arrive at the somewhat surprising
conclusion that the linear moments of the particle density,
as well as the linear momentum, must vanish even in the presence
of off-center vortices which lack axial or reflection symmetry.
Illustrations are provided by some analytical results in the limit
of a dilute gas, and by a numerical calculation of a class of single
and double vortices at intermediate couplings. The effect of anharmonic
confinement is also discussed.
\end{abstract}

\maketitle


Quantized vortices observed in a bulk superfluid such as $^4$He
have fascinated physicists for a long time \cite{donnelly}
because they provide definite macroscopic manifestations of
subtle quantum phenomena.
The subject has been significantly enriched in recent years with
      the realization of ultracold atomic Bose-Einstein condensates (BECs)
      confined in a finite region, where the strength of effective
      interactions may be manipulated by varying the number of atoms
      in a given trap. Typically, these condensates are sufficiently
      dilute that a mean-field Gross-Pitaevskii (GP) approximation
      is reliable. It is then possible to explicitly calculate a variety
      of vortex states which are relevant to experiment.
There have been numerous contributions
in this area, partly reviewed in Ref.~\cite{fetter}, but the subject is
still active because a number of finer issues remain unexplored.

The GP approximation is adopted throughout this paper.
We mainly consider an effectively two-dimensional (2D) Bose gas of $N$ atoms,
each with mass $M$, which interact pairwise with a contact potential
of positive strength $U_0$ and are confined by an axially symmetric
external potential $\hbar\omega_0 V(\rho/a_0)$.
Here the constant $\omega_0$ carries dimensions of frequency,
$a_0=\sqrt{\hbar/M\omega_0}$ is the corresponding oscillator length,
and $\rho=\sqrt{x^2+y^2}$ is the radial distance from the center of the trap.
Rationalized units are introduced by measuring time $t$ in units
of $1/\omega_0$, distances $x$ and $y$ in units of $a_0$,
while the condensate wave function is rescaled according to
$\Psi \to \sqrt{N}\,\Psi/a_0$ and thus acquires unit norm:
$\int{\Psi^*\Psi\, dx dy} =1$.
The energy functional is then given by
\begin{equation}  \label{eq:energy}
  E =  \int{\left[ \frac{1}{2}\,\left( \bm{\nabla}\Psi^* \bm{\nabla}\Psi \right)
 + V(\rho)\,\Psi^*\Psi + \frac{g}{2} (\Psi^*\Psi)^2 \right]\, dx dy},
\end{equation}
and yields energy in units of $\hbar\omega_0 N$; $V(\rho)$ is the
rationalized trap potential and $g=M N U_0/\hbar^2$ is a dimensionless
coupling constant.

In a frame rotating about the center of the trap with constant
angular frequency $\omega$ (in units of $\omega_0$) stationary states
of the gas satisfy the time-independent differential equations
\begin{eqnarray}  \label{eq:motion}
\mu \Psi - i \omega \eab x_\alpha \partial_\beta \Psi
  & = & \frac{\delta E}{\delta \Psi^*},  \nonumber \\
\mu \Psi^* + i \omega \eab x_\alpha \partial_\beta \Psi^*
  & = & \frac{\delta E}{\delta \Psi},
\end{eqnarray}
where
\begin{equation}  \label{eq:deltaE}
\frac{\delta E}{\delta \Psi^*} = -\frac{1}{2}\, \Delta \Psi
+ V(\rho) \Psi + g \, (\Psi^*\Psi) \Psi
\end{equation}
and $\delta E/\delta \Psi$ is its complex conjugate.
A chemical potential $\mu$ (in units of $\hbar\omega_0$) is introduced in
Eq.~(\ref{eq:motion}) in order to enforce a definite number of particles.
Greek indices $\alpha, \beta$ take two distinct values
corresponding to the two spatial coordinates $x_1=x$ and $x_2=y$,
and the Einstein summation convention of the repeated (dummy)
indices is consistently employed throughout the paper.
Finally, $\eab$ is the usual 2D antisymmetric tensor
and $\partial_\alpha=\partial/\partial x_\alpha$.

Now, given a solution $\Psi=\Psi(x,y | \mu,\omega)$ of Eqs.~(\ref{eq:motion}),
the time dependent wave function
\begin{eqnarray}  \label{eq:rotating}
\bar{\Psi}(x,y,t) & = & \Psi(x',y' | \mu,\omega)\, e^{-i \mu t}, \\
x' = x \cos\omega t & + & y \sin\omega t, \quad
y' = -x \sin\omega t + y \cos\omega t, \nonumber 
\end{eqnarray}
satisfies the standard Gross-Pitaevskii equation in the laboratory frame
($i\,\partial\bar{\Psi}/\partial t = \delta E/\delta \bar{\Psi}^*$) and may be
thought of as a configuration that rotates (precesses) about the center
of the trap with angular frequency $\omega$.
While all calculations will be based on the stationary Eqs.~(\ref{eq:motion}),
Eq.~(\ref{eq:rotating}) is important for a proper interpretation 
of the results.

Before discussing explicit solutions, we derive a class of virial theorems
which follow directly from Eqs.~(\ref{eq:motion}).
Thus we multiply both sides of the first equation by $\Psi^*$,
the second by $\Psi$, and add the two equations.
We then integrate both sides over the entire $xy$ plane and apply partial
integration to obtain
\begin{equation}  \label{eq:virial1}
\mu + \omega \ell = E_{\rm kin} + E_{\rm trap} + 2 E_{\rm pot},
\end{equation}
where $E_{\rm kin}$, $E_{\rm trap}$ and $E_{\rm pot}$ correspond to the
three terms in the total energy $E$ of Eq.~(\ref{eq:energy}).
In the left hand side of Eq.~(\ref{eq:virial1}) we have employed the
relations
\begin{equation}  \label{eq:angmom}
\int{\Psi^* \Psi\, dx dy} = 1, \quad
\ell = \frac{1}{i}\,\int{\Psi^* \eab
x_\alpha \partial_\beta \Psi\, dx dy},
\end{equation}
where the first is consistent with our choice of rationalized units
and the second is the definition of the angular momentum per particle
(in units of $\hbar$). Eq.~(\ref{eq:virial1}) is the first of a series
of virial relations that must be satisfied by all solutions of
Eqs.~(\ref{eq:motion}).

We now repeat the procedure by multiplying the first Eq.~(\ref{eq:motion})
by $\Psi^*$, the second by $\Psi$, and then subtracting the two equations
to obtain
\begin{eqnarray}  \label{eq:current}
& & -\omega\, \eab x_\alpha \d_\beta n + \d_\alpha J_\alpha = 0, \\
& & n = \Psi^* \Psi, \quad
J_\alpha = \frac{1}{2 i}\,
(\Psi^* \d_\alpha \Psi - \Psi \d_\alpha \Psi^*),  \nonumber
\end{eqnarray}
where $n$ and $\bm{J}$ are the familiar particle and current densities.
The same result could be derived by applying the continuity equation
in the laboratory frame  ($\d n/\d t + \bm{\nabla}\cdot\bm{J} = 0$)
for a wave function of the form (\ref{eq:rotating}).
Now multiply both sides of Eq.~(\ref{eq:current}) by $x_\alpha$ and then
apply partial integration
to obtain the virial relation
\begin{eqnarray}  \label{eq:virial2}
P_\alpha & + & \omega \eab R_\beta = 0,  \\
P_\alpha & \equiv & \int{J_\alpha\, dx dy}, \quad
R_\alpha \equiv \int{x_\alpha n\, dx dy},  \nonumber
\end{eqnarray}
where $\bm{P}=(P_1,P_2)$ is the linear momentum in the rotating frame
and $\bm{R}=(R_1,R_2)$ may be thought of as the mean position of the
configuration in question.
It should be clear that neither $\bm{P}$ nor $\bm{R}$ is conserved in the
laboratory frame, where they precess about the center with frequency
$\omega$, in complete analogy  with the momentum and position of a pointlike
particle in circular motion. In view of this analogy the virial relation
(\ref{eq:virial2}) appears to be quite natural.

A third and more elaborate class of virial relations is obtained starting
again from Eqs.~(\ref{eq:motion}) but now multiply the first equation
by $\d_\alpha \Psi^*$, the second by $\d_\alpha \Psi$, and then add the two
equations to yield after some rearrangement
\begin{equation}  \label{eq:mugamma}
\mu \d_\alpha n + \omega\, x_\alpha \gamma =
\frac{\delta E}{\delta \Psi^*}\, \d_\alpha \Psi^* +
\frac{\delta E}{\delta \Psi}\, \d_\alpha \Psi \equiv f_\alpha
\end{equation}
where $n= \Psi^* \Psi$ is again the particle density,
\begin{equation}  \label{eq:gamma}
 \gamma = \frac{1}{i}\, \eab\, \d_\alpha\Psi^* \d_\beta \Psi
\end{equation}
may be referred to as the topological vorticity \cite{pap1,pap2,komineas},
and we further use Eq.~(\ref{eq:deltaE})  to write
\begin{eqnarray}  \label{eq:stresstensor}
f_\alpha & = & \d_\beta \sigma_{\alpha\beta} + V(\rho) \d_\alpha n,
\nonumber \\
\sigma_{\alpha\beta} & = & w\, \delta_{\alpha\beta} - \frac{1}{2}\,
(\d_\alpha \Psi^* \d_\beta \Psi + \d_\beta \Psi^* \d_\alpha \Psi),  \\
w & = & \frac{1}{2} \, \left[ (\bm{\nabla}\Psi^*\cdot \bm{\nabla} \Psi)
 + g (\Psi^* \Psi)^2 \right].  \nonumber
\end{eqnarray}
Hence Eq.~(\ref{eq:mugamma}) reduces to a more transparent form:
\begin{equation}  \label{eq:mugamma2}
\mu \d_\alpha n + \omega\, x_\alpha \gamma =
\d_\beta \sigma_{\alpha\beta} + V(\rho) \d_\alpha n,
\end{equation}
which will provide the basis for the derivation of a number of
interesting virial relations.

Some key elements of the preceding discussion, such as the topological
vorticity $\gamma$ and the tensor $\sigma_{\alpha\beta}$, appeared
earlier in a study of the magnetic continuum \cite{pap1,komineas}
as well as of homogeneous superfluids \cite{pap2}.
In the latter case, the total topological vorticity
$\Gamma = \int{\gamma\, dx dy} = 2\pi n$
is integer valued ($n=0,\pm 1, \pm 2, \ldots$) for wave functions that
satisfy the boundary condition $|\Psi| \to 1$ at spatial infinity.
However, for the confined gas under present consideration,
the relevant wave functions satisfy the boundary condition $|\Psi| \to 0$
which leads to $\Gamma=0$ by a straightforward partial integration.
Similarly, the linear and angular momenta defined from
\begin{equation}  \label{eq:momenta}
P_\alpha = \int{\eab\,x_\beta\, \gamma \, dx dy}, \quad
\ell = -\frac{1}{2}\, \int{\rho^2\gamma\, dx dy},
\end{equation}
may be shown to coincide with the standard definitions given in
Eqs.~(\ref{eq:angmom}) and (\ref{eq:virial2}) by freely performing partial
integrations which are fully justified in a confined gas.
In contrast, partial integrations are generally ambiguous in a
homogeneous gas and Eqs.~(\ref{eq:momenta}) do not coincide with the
standard definitions, thus leading to a subtle distinction between
momentum and impulse \cite{batchelor,saffman}.
Certainly, for our current purposes, Eqs.~(\ref{eq:momenta}) may be employed
in conjunction with Eq.~(\ref{eq:mugamma2}) without further questioning.

Thus we integrate both sides of Eq.~(\ref{eq:mugamma2}) over the
entire $xy$ plane and note that the terms
$\d_\alpha n$ and $\d_\beta \sigma_{\alpha\beta}$
lead to vanishing surface integrals at spatial infinity.
The remaining terms may be arranged to yield the virial relation
\begin{equation}  \label{eq:virial3}
\omega P_\alpha + \eab\, \int{\frac{x_\beta}{\rho}\, V'\, n\, dx dy} = 0,
\end{equation}
where we have employed the linear momentum $P_\alpha$ from
Eq.~(\ref{eq:momenta}) and performed a partial integration in the second term,
with $V' = dV/d\rho$.

The basic relation (\ref{eq:mugamma2}) may be further iterated by
multiplying both sides with $x_\beta$ and then integrating over all
space to obtain
\begin{eqnarray}  \label{eq:15}
& &  \mu\, \delta_\ab - \omega \int{x_\alpha x_\beta\, \gamma\, dx dy} =
\nonumber \\
& & \int{\sigma_\ab \, dx dy} + \int{\left( V\delta_\ab +
\frac{x_\alpha x_\beta}{\rho} V'\right)\, n\, dx dy},
\end{eqnarray}
which may be applied for any combination of indices $\alpha$ and $\beta$
and thus contains three independent virial relations.
An interesting special case is obtained by taking the trace of both
sides of Eq.~(\ref{eq:15}):
\begin{eqnarray}  \label{eq:trace}
& & 2 \mu - \omega\, \int{\rho^2\gamma\, dx dy} =  \nonumber \\
& & \int{{\rm tr}\sigma\, dx dy} +
\int{(2V + \rho V')\, n\, dx dy},
\end{eqnarray}
where we may further insert the definition of the angular momentum
$\ell$ from Eq.~(\ref{eq:momenta}) and $tr\sigma=g(\Psi^*\Psi)^2$
from Eq.~(\ref{eq:stresstensor}) to write
\begin{equation}  \label{eq:virial4}
\mu + \omega \ell =
E_{\rm pot} + \int{(V + \frac{1}{2}\, \rho\, V')\,n\, dx dy}.
\end{equation}
This virial relation may be derived also by applying a Derrick-like
\cite{derrick} scaling argument to the extended energy functional
$F = E - \omega L - \mu N$.

Eqs.~(\ref{eq:virial1}), (\ref{eq:virial2}),
(\ref{eq:virial3}) and (\ref{eq:virial4}) already provide an interesting variety
of virial theorems which are employed in the following to check
and analyze explicit solutions of Eqs.~(\ref{eq:motion}).
It should be noted that virial relations (\ref{eq:virial1})
and (\ref{eq:virial2}) are insensitive to the specific choice of the
trap potential, while (\ref{eq:virial3}) and (\ref{eq:virial4})
depend crucially on the choice of $V=V(\rho)$.
Similarly, Eqs.~(\ref{eq:virial2}) and (\ref{eq:virial3})
are insensitive to the specific
      form of the tensor $\sigma_{\alpha\beta}$ and are thus valid for any type
      of interparticle interactions.

Most of the theoretical models employed to describe realistic BECs
assume a harmonic trap potential $V=\frac{1}{2}\rho^2$ and, hence,
$V'=\rho$. The virial relation (\ref{eq:virial4}) may then be written in
the form $\mu + \omega \ell = E_{\rm pot} + 2 E_{\rm trap}$ which is
combined  with Eq.~(\ref{eq:virial1}) to yield
\begin{equation}  \label{eq:virial5}
E_{\rm kin} + E_{\rm pot} = E_{\rm trap}.
\end{equation}
This relation does not contradict the existence of finite-energy
stationary solutions in the rotating frame and must indeed be verified by any
such solution of Eqs.~(\ref{eq:motion}). Similarly, the virial relation
(\ref{eq:virial3}) simplifies for $V'=\rho$ to read
$\omega P_\alpha + \eab R_\beta = 0$ which may be combined with
Eq.~(\ref{eq:virial2}) to arrive at the somewhat surprising conclusion
that both the linear momentum and the linear moments of the particle
density must vanish in a harmonic trap:
\begin{equation}  \label{eq:moments0}
P_\alpha = 0 = R_\alpha,
\end{equation}
provided that $\omega \neq 1$, a restriction that is not essential
      in the case of repulsive interactions because stationary states
      are then possible in a harmonic trap only
for $\omega < 1 (=\omega_0)$.
      In the special limit $\omega=1$, which could be achieved in the
    case of attractive interactions \cite{wilkin}, $P$ and $R$ need not vanish.

          A simple explanation of the preceding result can be obtained by
noting that for a system of atoms in
a harmonic trap with translationally-invariant interparticle interactions,
      the center-of-mass (CM)  coordinate separates from the internal
      coordinates and behaves as a free particle in a harmonic well.
    For $\omega<1$, the CM must be in its ground state, so $P$ and $R$ must be 
      zero because they depend only on the CM coordinates. At $\omega=1$,
      it is possible to put the CM into a rotating state with nonzero
      $P$ and $R$. This is relevant for the case of attractive interactions
      where the  rotating states with $\omega=1$ are such that the angular 
      momentum is carried by the center-of-mass \cite{wilkin}.
      This viewpoint
      makes it clear that the same result will apply for
      any translationally invariant interparticle interactions, in the
      case of 3D harmonic confinement, and beyond the GP
      approximation
(for the expectation value of the CM and conjugate momentum).
      It is also clear that Eq. (19) is not valid in the case of
      anharmonic confinement, as discussed later in this paper.

Explicit solutions of the Gross-Pitaevskii theory
were initially obtained in the limit of a very dilute
gas \cite{butts}.
A wave function with definite angular momentum $\ell$ may then be restricted
to the lowest Landau level (LLL):
\begin{equation}  \label{eq:LLL}
\Psi = \sum_{m \geq 0} c_m\Psi_m, \quad
\Psi_m = \frac{z^m e^{-|z|^2/2}}{\sqrt{m! \pi}},
\end{equation}
where the sum extends over nonnegative integer $m$ (for positive $\ell$)
and $z=x + i y$. The unknown coefficients $c_m$ are calculated by minimizing
the total energy $E=\ell + g/2\, \int{(\Psi^*\Psi)^2\, dx dy}$
under the constraints $\sum|c_m|^2=1$ and $\sum m |c_m|^2=\ell$.
This task was initially \cite{butts} carried out numerically to furnish
an impressive variety of vortex states leading up to a vortex lattice
for large $\ell$. For small $\ell$, some analytical results were obtained
\cite{kavoulakis} by a perturbative expansion of the coefficients
$c_m$ in powers of $\ell$ or $\bar{\ell}=1-\ell$.

These results may already be used to illustrate
the virial relations (\ref{eq:moments0}). We first consider the linear
moments of the particle density $R_\alpha$, with $\alpha=1$ or 2,
or their complex combination $R_1+ i R_2 = \int{\Psi^* z \Psi\, dx dy}$.
We may then insert the series representation (\ref{eq:LLL}) for the
wave function $\Psi$, note that $z\Psi_m = \sqrt{m+1} \Psi_{m+1}$, and
apply the usual orthogonality relations for the $\Psi_m$'s to obtain
\begin{equation}  \label{eq:position}
R_1 + i R_2 = \sum_{m \geq 0} \sqrt{m+1}\, c_{m+1}^* c_m.
\end{equation}
If we now use the perturbative expansions for the $c_m$'s from
Ref.~\cite{kavoulakis}, with due attention to phase (sign) conventions
\cite{kavoulakis2}, we find that $R_1+ i R_2 = 0$ order-by-order
in a consistent expansion in powers of $\ell$ or $\bar{\ell}=1-\ell$.
A similar calculation of the linear momentum yields $P_1+i P_2 = 0$,
thus confirming the validity of both virial relations
in Eq.~(\ref{eq:moments0}), as well as providing a nontrivial check
of consistency of the results of Ref.~\cite{kavoulakis}.

A more convincing demonstration is possible over the entire range
$0<\ell<1$ where a closed-form expression for the optimal LLL
wave function was recently achieved \cite{vorov}:
\begin{eqnarray}  \label{eq:LLLwavefunction}
& & \Psi = \frac{\ell^\frac{1}{4}}{\sqrt{\pi}}\, [(x-b) + i y]\,
e^{-\frac{1}{2}[(x-a)^2 - 2 i a y + y^2]}  \nonumber \\
& & a = (\sqrt{\ell}-\ell)^{1/2}, \quad  b = \frac{1-\ell}{a},
\end{eqnarray}
which describes an off-center vortex located on the $x$ axis, modulo
an overall azimuthal rotation, at a distance  $b=b(\ell)$ from the
center of the trap. For $\ell \to 0\; (b \to \infty)$ the vortex is expelled
from the system, while for $\ell \to 1\;  (b \to 0)$ the vortex moves
to the center and becomes axially symmetric.
The corresponding particle density reads
\begin{equation}  \label{LLLdensity}
n = \frac{\sqrt{\ell}}{\pi}\, [ (x-b)^2 + y^2]\, e^{-(x-a)^2 - y^2}
\end{equation}
and lacks axial or reflection symmetry except for $\ell=0$ or 1.
Nevertheless, an explicit calculation shows that
\begin{equation}  \label{LLLposition}
R_1 = \int{x\, n\, dx dy} = \sqrt{\ell}\,[2 a - b + a (a-b)^2] = 0,
\end{equation}
using the explicit expressions for $a$ and $b$
from Eq.~(\ref{eq:LLLwavefunction}), and $R_2=\int{y n\, dx dy} = 0$,
thanks to the $y \to -y$ symmetry of the particle density.
A similar calculation shows that the linear momentum
$\bm{P} = (P_1,P_2)$ also vanishes,
thus verifying both virial relations in Eq.~(\ref{eq:moments0})
in spite of the lack of axial or reflection symmetry in the wave function
(\ref{eq:LLLwavefunction}).
This result appears to be surprising if we naively view the off-center
vortex as a pointlike particle rotating on a circle with radius $b$.
In fact, the spatial distribution of particle and current densities
is more subtle in an off-center vortex and leads to vanishing $\bm{P}$
and $\bm{R}$, even though the vortex does precess about the center of
the trap. As we shall see shortly, this curious result is valid only
for harmonic confinement ($V=\frac{1}{2}\rho^2$).

We complete the discussion of the dilute-gas limit by quoting the energy
$E$ and frequency $\omega$ associated with the wave function
(\ref{eq:LLLwavefunction}):
\begin{equation}  \label{eq:LLLenergy}
E = \ell + \frac{g}{4\pi}\, \left(1-\frac{\ell}{2}\right), \quad
\omega = \frac{dE}{d\ell} = 1 -\delta,
\end{equation}
with $\delta \equiv g/8 \pi$, a result that is valid in the limit $\delta \ll 1$.
A notable feature of this limit is that frequency $\omega$ is
independent of angular momentum $\ell$ \cite{wilkin,butts,kavoulakis,vorov}.

One should keep in mind that practically all experiments have been
performed on BECs with $\delta > 1$, often $\delta \gg 1$, where the
weak coupling (LLL) theory is no longer valid.
A numerical solution of Eqs.~(\ref{eq:motion}) is necessary for
strong couplings. A numerical method developed in
\cite{castin} is based on a norm-preserving relaxation algorithm
which, in effect, capitalizes on the virial relation ({\ref{eq:virial1}) to find
a wave function of unit norm that is a (local) minimum of the energy
functional in the rotating frame: $E_{\rm rot} = E - \omega \ell$.
Here we employ a variation of the norm-preserving algorithm to minimize
instead the Lyapunov functional $E' = E + \frac{1}{2} a (\ell-b)^2$
where $a$ and $b$ are arbitrary constants with $a>0$.
Local minima of $E'$ satisfy Eqs.~(\ref{eq:motion}) with frequency
determined self-consistently from $\omega=a (b-\ell)$ and chemical
potential $\mu$ from Eq.~(\ref{eq:virial1}).
The constants $a$ and $b$ are chosen (tuned) to ensure convergence
to nontrivial solutions with angular momentum $\ell$ in the desired range.
The advantage of this algorithm is that it finds solutions
of Eqs.~(\ref{eq:motion}) which are stationary points but not
necessarily local minima of the functional $E_{\rm rot}$.

\begin{figure}
\epsfig{file=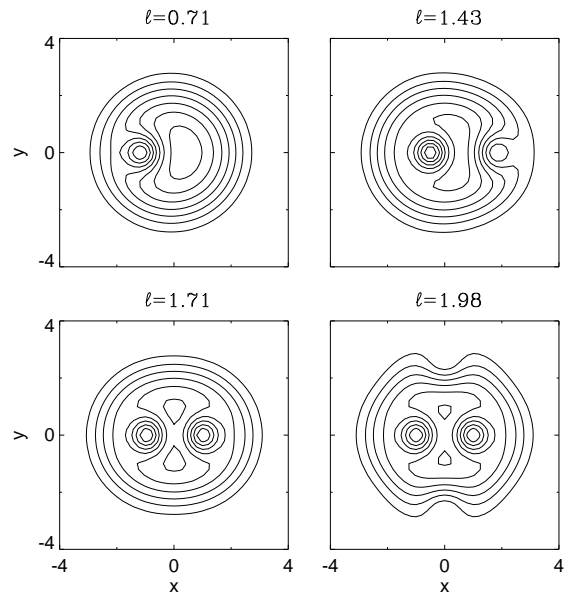,width=8cm}
 \caption{Contour plots of the particle density $n$ for a class
of vortex states calculated in a harmonic trap with $\delta = 2$.
The four characteristic values of the angular momentum $\ell$
correspond to the four branches $AB, B'\Gamma, \Gamma\Delta$, and $\Delta\ldots$
in Fig.~\ref{fig:harmonic}.}
  \label{fig:contour}
\end{figure}

\begin{figure}
\epsfig{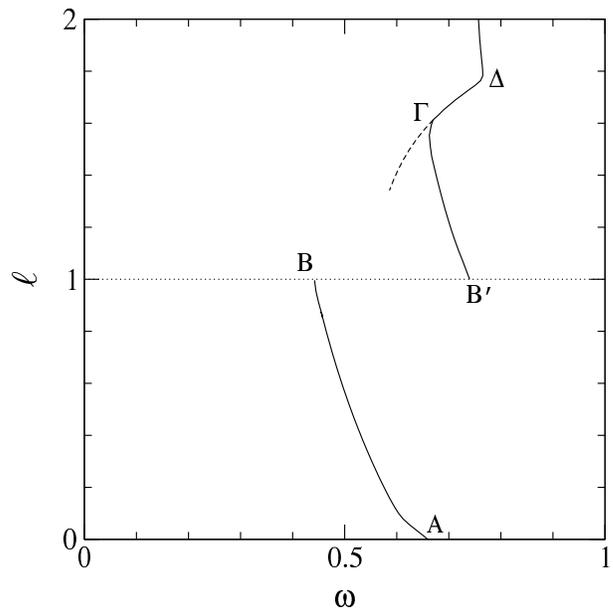}
 \caption{Angular frequency $\omega$ (in units of $\omega_0$)
as a function of angular momentum per particle $\ell$
(in units of $\hbar$) for a class of vortex states calculated in a harmonic
trap with $\delta=2$.
}
 \label{fig:harmonic}
\end{figure}

In the following we describe a class of solutions for the intermediate
coupling $\delta = g/8 \pi = 2$ where the LLL approximation is
quantitatively inaccurate.
Needless to say, all virial relations, including Eqs.~(\ref{eq:virial5})
and (\ref{eq:moments0}) were confirmed by our solutions to within
numerical accuracy. In Fig.~\ref{fig:contour} we present contour plots
of the particle density $n$ for four characteristic values of angular
momentum in the range $0 < \ell < 2$, whereas Fig.~\ref{fig:harmonic}
shows the results for the frequency dispersion
$\omega = dE/d\ell = \omega(\ell)$ throughout the same range.

For $0 < \ell < 1$ the calculated configuration is an off-center vortex
with energy $E(\ell)$ that is a concave function of $\ell$.
Thus the frequency $\omega = dE/d\ell$ is a decreasing function of
angular momentum taking values in the finite range
$\omega_B < \omega < \omega_A$ with $\omega_B = 0.44$ for $\ell \to 1^-$
and $\omega_A = 0.62$ for  $\ell \to 0$.
The vortex precesses faster the farther it is located form the center
of the trap ($\ell \to 0$). In the opposite limit ($\ell \to 1^-$)
the vortex moves to the center and becomes axially symmetric.
There is no sense of precession in such a vortex because the limiting
frequency $\omega_B$ may then be absorbed into an effective chemical potential
$\bar{\mu} = \mu + \omega_B$ and the wave function (\ref{eq:rotating})
reduces to a quasi-static configuration with chemical potential $\bar{\mu}$.
Also note that the band of allowed frequencies reduces to a single point
$\omega_A = \omega_B = 1 -\delta$ in the dilute-gas limit ($\delta \ll 1$),
as is evident from Eq.~(\ref{eq:LLLenergy}).

As the angular momentum increases beyond unity the energy remains continuous,
but its first derivative exhibits a finite jump which leads to a new
limiting frequency  $\omega_{B'} = 0.74$ for $\ell \to 1^+$. The original
vortex becomes again displaced from the center for $\ell = 1^+$ and a
second off-center vortex appears at an asymmetric position on the
opposite side of the trap.
This picture remains largely correct in the region $1 < \ell < 1.55$
and leads to branch $B'\Gamma$ in Fig.~\ref{fig:harmonic} with limiting
frequencies $\omega_{B'} = 0.74$ and $\omega_{\Gamma} = 0.66$.
Energy is again concave in this region and thus frequency is a decreasing
function of angular momentum.

At point $\Gamma\, (\ell=1.55)$ the calculated configuration becomes a
reflection-symmetric two-vortex state where the two vortices are located
at the same distance on opposite sides from the center of the trap.
Such a symmetric state persists throughout the branch $\Gamma\Delta$
($1.55 < \ell < 1.78$) with corresponding frequencies in the range
$0.66 < \omega < 0.76$.
Incidentally, $\Gamma\Delta$ is the only branch where the energy
$E(\ell)$ is convex and thus the frequency $\omega(\ell)$ is an increasing
function of angular momentum.

Beyond point $\Delta\, (\ell > 1.78)$ the frequency becomes once again
a decreasing function of angular momentum.
This region seems to be characterized by the appearance of a new pair
of vortices symmetrically displayed along the $y$ axis,
as indicated by the fourth ($\ell$=1.98) entry of Fig.~\ref{fig:contour}.
Nevertheless, reflection symmetry appears to persist in this region.
A related interesting question is whether or not hysteresis sets in when
we reverse the cycle by reducing the angular momentum from, say, $\ell=1.98$.
While the cycle is perfectly reproduced down to point $\Gamma$, a reflection
symmetric two-vortex state persists for some range of angular momenta
below $\ell_\Gamma=1.55$, as indicated by the dashed line
in Fig.~\ref{fig:harmonic}.

The issue of stability of the calculated vortex states is rather delicate
and may well depend on the specific experimental protocol.
According to \cite{butts} mechanical stability requires
that the energy be a convex function of angular momentum:
$d\omega/d\ell = d^2E/d\ell^2 > 0$.
This condition is satisfied only by the $\Gamma\Delta$ branch
of Fig.~\ref{fig:harmonic} which corresponds to symmetric two-vortex states.
In particular, the whole of the AB branch, which corresponds to single
off-center vortices, does not satisfy the criterion of mechanical stability.
Nevertheless, precessing off-center vortices have been observed
experimentally in a spherical trap \cite{anderson}.
Although the current two-dimensional calculation does not directly
apply to a spherical trap, a similar three-dimensional calculation
\cite{SUvortex} leads to a class of U-shaped off-center vortices 
whose frequency dispersion is completely analogous to the AB branch
of Fig.~\ref{fig:harmonic}.
Furthermore, the frequencies of precession measured in the experiment
lie within the calculated frequency band $[\omega_B, \omega_A]$.
To conclude this digression, we note that the virial relations
(\ref{eq:moments0}) are valid also in a three-dimensional axially
symmetric harmonic trap, with $\alpha=1$ or 2 corresponding to the directions
perpendicular to the symmetry (rotation) axis,
to be completed with $P_3=0=R_3$ along the same axis.

Finally, we consider the effect of anharmonic confinement
modelled here by the rationalized trap potential
\begin{equation}  \label{eq:anharmonic}
V = \frac{1}{2}\, \rho^2 (1 + \lambda \rho^2), \quad
V' = \rho + 2 \lambda \rho^3,
\end{equation}
which is thought to describe the trap used in the experiment of
Ref.~\cite{bretin} with $\lambda \sim 10^{-3}$.
In our numerical calculation we adopted a much larger $\lambda$
in order to emphasize some generic features of anharmonicity.
We have thus repeated our earlier calculation of vortex states in a
harmonic trap ($\lambda=0$) now for $\lambda=1/4$ but the same coupling
constant $\delta= g/8 \pi = 2$.

\begin{figure}
\epsfig{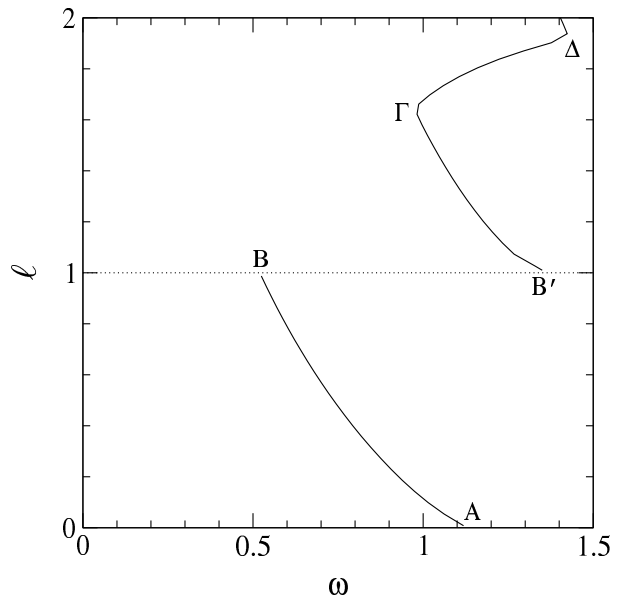}
 \caption{Angular frequency $\omega$
as a function of angular momentum per particle $\ell$
for a class of vortex states calculated in an anharmonic
trap with $\delta=2$ and $\lambda=1/4$.
}
 \label{fig:anharmonic}
\end{figure}

The calculated frequency dispersion is shown in Fig.~\ref{fig:anharmonic}
which differs from Fig.~\ref{fig:harmonic} mainly by the fact
that the dispersion now extends well beyond $\omega=1 (=\omega_0)$
because the condition $\omega < 1$ is no longer necessary to ensure
stability of the rotating gas. At first sight, the vortex configurations
that correspond to the various branches of Fig.~\ref{fig:anharmonic}
are also similar to those calculated for the harmonic trap and shown
in Fig.~\ref{fig:contour}. However, a closer look reveals some
subtle differences which are best illustrated by recalling the virial
relation (\ref{eq:virial3}) now applied for the potential $V$
of Eq.~(\ref{eq:anharmonic}):
\begin{eqnarray}  \label{eq:27}
& & \omega P_\alpha + \eab R_\beta = -2 \lambda \eab Q_\beta, \\
& & Q_\alpha \equiv \int{x_\alpha \rho^2 n \, dx dy}, \nonumber
\end{eqnarray}
where $Q_\alpha$ is a generalized (higher) moment of the particle density.
We further recall virial relation (\ref{eq:virial2}), which is valid for any
trap potential, and combine it with Eq.~(\ref{eq:27}) to yield
\begin{equation}  \label{eq:moments}
P_\alpha = \frac{2 \lambda \omega}{1-\omega^2}\, \eab Q_\beta, \quad
R_\alpha = -\frac{2 \lambda}{1-\omega^2}\,Q_\alpha,
\end{equation}
which differ significantly from Eqs.~(\ref{eq:moments0}) in that
the linear momentum $\bm{P}$ and moment $\bm{R}$ are no longer forced
to vanish.

\begin{figure}
\epsfig{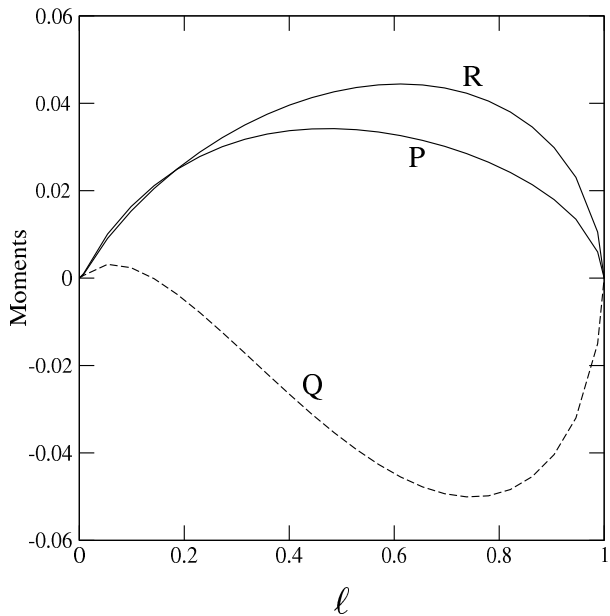}
 \caption{Linear momentum $P$ (in units of $\hbar/a_0$),
linear moment $R$ (in units of $a_0$), and generalized moment
$Q$ (in units of $a_0^3$), as functions of angular momentum
per particle $\ell$ (in units of $\hbar$), for single off-center
vortices in an anharmonic trap with $\delta=2$ and $\lambda=1/4$.
}
 \label{fig:moments}
\end{figure}

For a numerical illustration we consider the class of single off-center
vortices, which correspond to the AB branch of Fig.~\ref{fig:anharmonic},
and assume without loss of generality that the vortex is located on the
$x$ axis. We may then insert $\bm{P}=(0,P), \bm{R}=(R,0)$ and $\bm{Q}=(Q,0)$
in Eqs.~(\ref{eq:moments}) to obtain the more transparent relations
\begin{equation}  \label{eq:moments2}
P = -\frac{2 \lambda \omega}{1-\omega^2}\, Q, \quad
R = -\frac{2 \lambda}{1-\omega^2}\,Q.
\end{equation}
The moments $P,R$ and $Q$ calculated along the AB branch ($0 < \ell < 1$)
were found to satisfy the virial relations (\ref{eq:moments2}) and are
depicted as functions of angular momentum in Fig.~\ref{fig:moments}.
A notable fact is that all moments vanish for $\ell=0$ or 1
because the calculated  wave function becomes axially symmetric in both of
the above limits.
Nevertheless, the moments do not vanish for other values of the angular
momentum in the interval $0 < \ell < 1$. Viewed from the laboratory frame,
the linear momentum and mean position of the vortex read
\begin{eqnarray}
\bm{P}_{\rm lab} & = & P(-\sin\omega t, \cos\omega t), \nonumber \\
\bm{R}_{\rm lab} & = & R(\cos\omega t, \sin\omega t),
\end{eqnarray}
with $P=\omega R$, in complete analogy with the motion of a pointlike particle
rotating around the center. Therefore, generic behavior prevails
in the presence of some anharmonicity ($\lambda \neq 0$), whereas the
stronger virial relations (\ref{eq:moments0}) are but a curious
feature of the harmonic limit ($\lambda = 0$).

In conclusion, the specific family of solutions analyzed in this paper
illustrates some of the subtleties of vortex states in a confined
Bose-Einstein condensate but certainly does not exhaust the possibilities.
It is clear that a huge variety of multiple vortex states are possible,
with increasing angular momentum, which eventually lead to a formation
of regular vortex lattices. The virial theorems derived here must
be satisfied in all cases and may thus be used to provide important
checks of consistency, especially because they are sensitive to the
presence of anharmonicity.

\vspace{10pt}
SK and NRC are grateful to the Kavli Institute of Theoretical Physics
in Santa Barbara for hospitality and acknowledge discussions
during the ``Quantum gases'' program from which this work has benefited.
This work was supported by
EPSRC Grant Nos GR/R96026/01 (SK) and GR/S61263/01 (NRC).


\end{document}